\documentclass[twocolumn,pra,aps,showpacs,floatfix]{revtex4}
\usepackage{url,graphicx,amssymb,amsmath}

\newcommand{\br}{\boldsymbol r}

\begin{document}

\title{Time-dependent electron localization function} 

\author{T. Burnus}
\author{M\,A.\,L. Marques}
\author{E\,K.\,U. Gross}
\email{hardy@physik.fu-berlin.de}
\affiliation{Institut f\"ur Theoretische Physik, Freie Universit\"at Berlin, Arnimallee 14, 14195 Berlin, Germany}

\date{\today}

\begin{abstract}
We present a generalization of the electron localization
function (ELF) that can be used to analyze time-dependent processes. The time-dependent ELF
allows the {\it time-resolved} observation of the formation, the modulation, and the 
breaking of chemical bonds, and can thus provide a visual understanding of complex reactions 
involving the dynamics of excited electrons. We illustrate the usefulness of the time-dependent 
ELF by two examples: the $\pi$-$\pi^*$ transition induced by a laser field, and 
the destruction of bonds and formation of lone pairs in a scattering process.
\end{abstract}


\maketitle

The advent of ultrashort laser sources with pulse durations on the order of 10--100 femtoseconds \cite{nlo}
has paved the way to the analysis and control of chemical reactions \cite{pZuwail2000}: By means of
pump-probe spectroscopies with femtosecond laser pulses \cite{pZuwail1995} one can follow, in the
time domain, the nuclear motion which typically evolves on the picosecond time scale. One of
the most important recent achievements has been the experimental realization of attosecond
pulses\cite{pPaul2001}. These are produced by coherently superimposing high harmonics
generated by a strong infrared laser pulse interacting with atoms. With this light source
available, pump-probe spectroscopies using attosecond pulses allow the 
temporal resolution of the electronic motion which is governed by the femtosecond time scale.
Questions like ``How does an electron travel from the highest occupied to the lowest unoccupied molecular orbital
when excited by a laser?'' may soon become experimentally accessible. This 
establishes the need for theoretical tools to analyze and interpret such data. Theoretical tools
of this type will be developed in this article. We are mainly concerned with bonds, and how they 
break, form, or change during a time-dependent process.

The intuitive concept of a chemical bond is very simple and elegant: 
an electron pair shared between neighboring atoms 
that provides the necessary attraction to bind the molecule. However, it turns out to be very difficult 
to define exactly what a bond is, or even to visualize it. The one-electron molecular orbitals that 
stem from density-functional theory or Hartree-Fock usually have contributions from several atoms 
and do not represent a unique bond. The electronic density, on the other hand, does not easily
reveal important features like lone pairs. 
The electron localization function (ELF) is a function crafted to bring into evidence the subtle 
bonding properties of an electronic system. 
It was originally applied to ground-state systems, in the study of atomic shells and 
covalent bonds \cite{Becke1990}. Soon after, it was realized that the ELF could be also used to 
analyze lone pairs, hydrogen bonds \cite{Chesnut2000}, 
ionic and metallic bonds \cite{Savin1997}, etc. The systems studied include atoms \cite{Becke1990}, 
molecules\cite{Savin1997}, surfaces \cite{Savin1997} and solids \cite{Burdett1998,Tsirelson2002,Savin1997}. 
It is also possible to establish a rigorous topological classification of chemical
bonds using the ELF \cite{Silvi1994}. Furthermore, the ELF has the 
advantage of being fairly insensitive to the method used to calculate the wave functions of the 
system. In fact, Hartree-Fock, density-functional theory, or even simple approaches such as extended 
H\"uckel methods, yield qualitatively similar ELFs \cite{Savin1997}. Approximate electron 
localization functions have also been obtained from experimental electron densities measured 
with x-ray \cite{Tsirelson2002}.

\begin{figure*}[t]
\begin{center}
  \includegraphics*[scale=0.95]{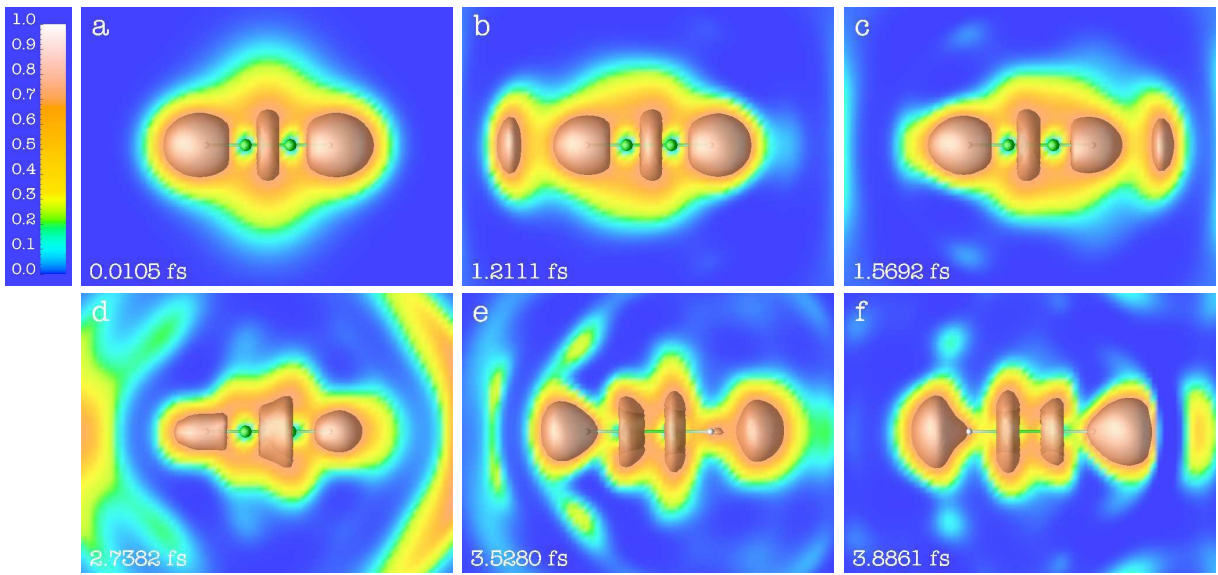}
\end{center}
\caption{(Color) Snapshots of the time-dependent ELF for the 
excitation of acetylene by a 17.15~eV laser pulse \cite{movie}.
The pulse had a total length of 7~fs, an intensity of $1.2\times10^{14}$~W\,cm$^{-2}$, and was 
polarized along the molecular axis. Ionization and the transition from the bonding $\pi$ state
to the antibonding $\pi^*$ state are clearly visible.
\label{fig1}}
\end{figure*}

Up to now the ELF has only been used to study systems in their ground state. Clearly, an 
extension of the ELF to time-dependent processes appears quite desirable. Such an extension
would allow the {\it time-resolved} observation of the formation, the modulation, and the 
breaking of chemical bonds, and thus provide a visual understanding of complex reactions 
involving the dynamics of excited electrons. That is exactly the goal of this communication.

The derivation of a time-dependent ELF (TDELF) follows closely the derivation of Becke and Edgecombe 
of the ground-state ELF \cite{Becke1990}.
Our objective is to find a function, $D_\sigma(\br, t)$, that measures the probability of finding one
electron in the near vicinity of a reference like-spin electron at position $\br$ and time $t$. 
If this probability is high then the reference electron must be delocalized. On the other hand, 
a covalent bond consists of a localized electron pair (of opposite spins) between two 
neighboring atoms. The probability of finding a like-spin electron close to a reference electron 
in this region must then be very low. For a determinantal many-body wave function, the
function $D_\sigma(\br, t)$ is given by
\begin{equation}
  \label{Dsigma}
  D_\sigma(\br, t) = \tau_\sigma(\br, t)
    - \frac{1}{4}\frac{\bigl[\nabla n_\sigma(\br,t)\bigr]^2}{n_\sigma(\br, t)}
    - \frac{j^2_\sigma(\br, t)}{n_\sigma(\br, t)},
\end{equation}
where $\sigma$ denotes the spin, $n_\sigma$ the spin density, $j_\sigma$ the absolute value
of the current density, and
\begin{equation}
  \tau_\sigma(\br, t) = \sum_{i=1}^{N_\sigma} \left|\nabla \varphi_{i\sigma}(\br, t)\right|^2.
\end{equation}
This last expression represents the kinetic-energy density of a system of $N_\sigma$ electrons,
described by the one-particle orbitals $\varphi_{i\sigma}$. These orbitals can be obtained, e.g.,
from time-dependent density-functional theory or from a time-dependent Hartree-Fock calculation.

Equation~(\ref{Dsigma}) is similar to the expression for the ground-state 
$D_\sigma(\br)$ \cite{Becke1990}. The main difference is the additional term
proportional to $j_\sigma^2$ \cite{pDobson1993}. This term naturally arises when the analysis of
Ref.~[\onlinecite{Becke1990}] is carried out without assuming real-valued orbitals, i.e.,
without assigning vanishing orbital currents. This was pointed out by Dobson \cite{pDobson1993}
in his evaluation of the Fermi-hole curvature of a static current-carrying single-determinant state.
It is easy to see that the derivation of Ref.~[\onlinecite{pDobson1993}] applies equally well to
time-dependent determinantal wave functions, thus leading to Eq.~(\ref{Dsigma}).

The function $D_\sigma(\br, t)$ is always $\ge 0$, but it is not bounded from above. As
usual, we define as an alternative measure of localization
\begin{equation}
 f_\mathrm{ELF}(\br, t) = \frac{1}{1+\bigl[D_\sigma(\br, t)/D_\sigma^0(\br, t)\bigr]^2},
\end{equation}
with the definition $D_\sigma^0(\br, t) = \tau_\sigma^\mathrm{HEG}\bigl(n_\sigma(\br,t)\bigr)$, where
\begin{equation}
  \tau_\sigma^\mathrm{HEG}(n_\sigma) = \frac{3}{5}(6\pi^2)^{2/3}
    n_\sigma^{5/3}
\end{equation}
is the kinetic-energy density of a homogeneous electron gas of density $n_\sigma$. Using this 
definition, $f_\mathrm{ELF}$ is dimensionless and lies between zero and one. 
A value of 1, i.e., $D_\sigma(\br, t)$ approaching zero, corresponds to high localization.

\begin{figure*}[t]
\begin{center}
  \includegraphics*[scale=1.0]{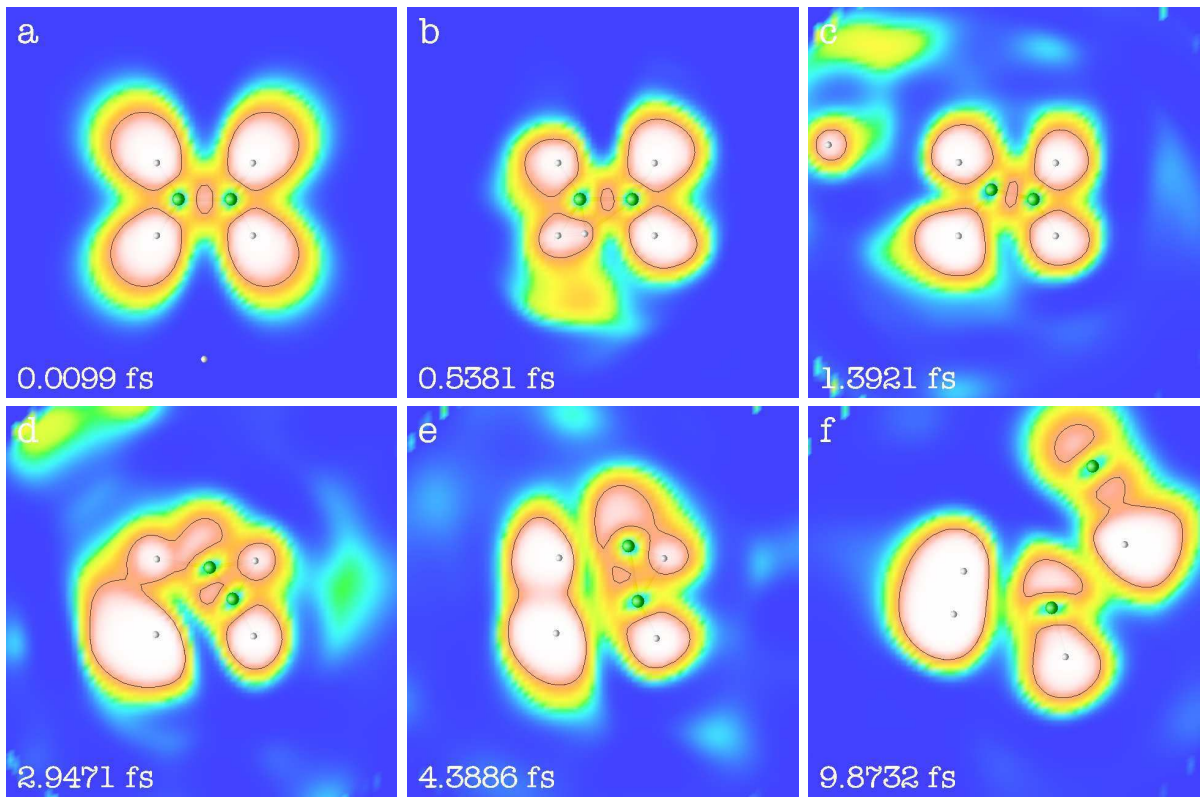}
\end{center}
\caption{(Color)
Snapshots of the time-dependent ELF for the scattering of a fast, nonrelativistic proton 
($E_{\rm kin}\approx 2$~keV) by ethene \cite{movie}. The molecule breaks into several pieces.
During this fragmentation process, the breaking of bonds and the subsequent creation of several 
lone pairs become clearly visible in the time-dependent ELF. 
The legend is the same as in Fig.~\ref{fig1}.
\label{fig2}}
\end{figure*}

We illustrate the usefulness of the time-dependent ELF by two examples: 
(i) the excitation of acetylene by a strong laser pulse (Fig.~\ref{fig1}), and 
(ii) the scattering of a high-energy proton from the ethene molecule (Fig.~\ref{fig2}).
The figures depict a slab of the ELF passing through the plane of the molecules, to which
we superimposed an isosurface (Fig.~\ref{fig1}) or contour lines
(Fig.~\ref{fig2}) at $f_\mathrm{ELF}=0.8$. 
Movies of the time-dependent ELF and of the corresponding time-dependent density
can be found in our website \cite{website} or as supplementary material to this article \cite{movie}. 
All calculations were performed in the 
framework of time-dependent density-functional theory \cite{runge1984,pMarques2003},
using a real-space, real-time code \cite{octopus}, and employing the adiabatic local-density
approximation \cite{pMarques2003} for the time-dependent exchange-correlation
potential. This approximation is expected to overestimate ionization due to
spurious self-interactions. This problem can be overcome by the use of explicitly
orbital-dependent functionals within the time-dependent optimized effective
potential method \cite{pUllrich1995}.

In the beginning of the simulation (i) the acetylene molecule is in its ground state
[Fig.~\ref{fig1}(a)]. At this moment, the ELF exhibits three
major features: a torus between the carbon atoms---typical of triple
bonds---and the two characteristic blobs around the hydrogens. As the intensity of
the laser increases, the system starts to oscillate, and then to ionize
[Figs.~\ref{fig1}(b) and \ref{fig1}(c)]. Note that the ionized charge leaves the
system in fairly localized packets [the blob on the left in panel (b), and on the
right in panel (c)], that then spread with time. The central torus then starts to 
widen until the moment it breaks into two separate tori, each around one carbon atom
[Fig.~\ref{fig1}(e)]. We interpret this finding as a transition
from the $\pi$ bonding state into the $\pi^*$ nonbonding state. The system then remains 
in this excited state for some time. We emphasize that our calculation corresponds to
one specific orientation of the molecule, namely the orientation where the polarization
vector of the laser field is parallel to the molecular axis. A detailed analysis of the 
dependence of photoionization on the molecular orientation will be presented 
elsewhere \cite{pCastroU}.
 
\begin{figure}
%\begin{center}
  \includegraphics[scale=1.0]{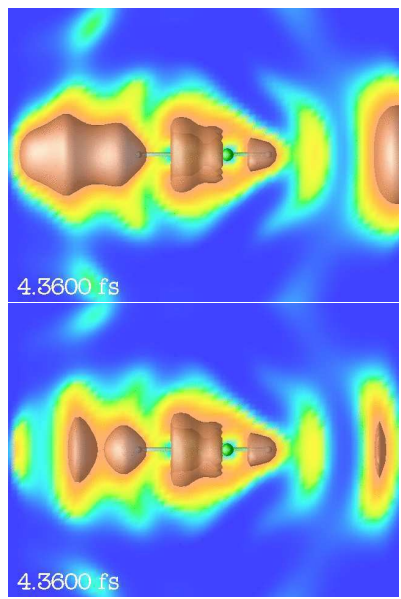}
%\end{center}
\caption{(Color)
Snapshot of the time-dependent ELF with (upper panel) and without (lower panel) the
$j^2/n$ term for the excitation of acetylene by a 17.15~eV laser pulse.
The legend is the same as in Fig.~\ref{fig1}.
\label{fig3}}
\end{figure}

To assess the importance of the $j^2/n$ term, we compare, in Fig.~\ref{fig3}, a snapshot
of the proper time-dependent ELF [given by Eq.~(\ref{Dsigma})]  with the traditional
ground-state expression (where the $j^2/n$ term is absent). The plot shows clear 
differences in some regions of space, thus underlining the significance of the 
$j^2/n$ term.

In our second example we send a fast, but still nonrelativistic, proton against one
of the carbons of ethene (Fig.~\ref{fig2}). The initial configuration
is shown in panel (a). While approaching the carbon atom the proton accumulates
some charge around it [Fig.~\ref{fig2}(b)]. It then scatters 
and leaves the system picking up part of the electronic charge. The electron-nuclei system is 
thus excited (in total the electronic system absorbs around 20~eV). 
In panels (d) and (e) the leftmost carbon has already broken the two bonds with
the hydrogens (that will later form an H$_2$ molecule). Clearly visible is also
the polarization of the carbon-carbon double bond, and the starting of the formation
of a lone pair above the leftmost carbon. We emphasize once again that the formation
of lone pairs cannot be visualized in movies of the time-dependent density. Only the
TDELF allows one to observe this feature.
At the end of the simulation [panel (f)]
we can observe an H$_2$ molecule (left), and two CH fragments (middle and right). 
The rightmost CH fragment is again breaking to yield a carbon and a hydrogen atom. 
Note again the lone pairs characteristic of CH, localized near the 
carbon atoms.

These two examples illustrate the amount of information
that can immediately be grasped just by looking at the time-dependent ELF,
from $\pi$-$\pi^*$ transitions, to the creation of lone pairs.
One can infer the {\it time scale} and the {\it temporal order} of
the various processes occurring, e.g., in a molecular collision: One
can tell which bond breaks first and which breaks second, and how many attoseconds or femtoseconds
it takes to form new bonds and new lone pairs. We emphasize once more that with the advent 
of attosecond pulses this information will soon become experimentally available.
We expect the time-dependent ELF to be a valuable tool in the analysis of many
physical processes. One example is the creation and decay of collective 
excitations or the scattering of electrons from atoms and molecules. 
Another example is the process of vision: light promotes the electrons of 
retinal into the first excited state, which by its turn induces the isomerization 
of the photo-receptor and eventually leads to the firing of a neuron.
The key feature of the time-dependent ELF, in all cases, is the {\it time-resolved}
observation of the formation, the modulation, or the breaking of chemical bonds, thus
providing a visual understanding of the dynamics of excited electrons.

%\section*{Acknowledgements}
This work was supported in part by the NANOQUANTA Network of Excellence, by the EXC!TiNG
Research and Training Network and by the Deutsche Forschungsgemeinschaft within the
Sonderforschungsbereich SFB450.

\bibliographystyle{Science}

\bibliography{scibib}

\end{document}